\documentclass{article}
\usepackage{spconfa4,amsmath,graphicx}
\usepackage{lipsum}
\usepackage{amssymb}
\usepackage{algorithm}
\usepackage{algpseudocode}
\usepackage{color}
\usepackage{xcolor}
\usepackage{booktabs}
\usepackage{lipsum}
\usepackage{svg}
\usepackage{varwidth}
\usepackage{tikz}
\usepackage{pgf}
\usepackage{pgfplots}
\usepackage{pgfplotstable}
\usetikzlibrary{positioning,backgrounds,fit,calc,patterns,arrows.meta}
\pgfplotsset{compat=1.3}
\usetikzlibrary{shapes.geometric}
\usetikzlibrary{shapes.arrows}
\usetikzlibrary{arrows.meta}
\usetikzlibrary{external}
\usepackage[skins]{tcolorbox}
\usepackage{adjustbox}
\usepackage{hyperref}

\title{Gaussian Flow Bridges for Audio Domain Transfer with Unpaired Data}
\twoauthors
  {Eloi Moliner\sthanks{Work done while intern at Microsoft Research}\vspace{-10pt}}
	{Acoustics Lab, DICE \\
	Aalto University, Espoo, Finland \\
eloi.moliner@aalto.fi }
  {Sebastian Braun \quad Hannes Gamper}
	{Microsoft Research Redmond\\
	{first.last}@microsoft.com}
\begin{document}
\ninept
\maketitle
\begin{abstract}
Audio domain transfer is the process of modifying audio signals to match characteristics of a different domain, while retaining the original content. This paper investigates the potential of Gaussian Flow Bridges, an emerging approach in generative modeling, for this problem. The presented framework addresses the transport problem across different distributions of audio signals through the implementation of a series of two deterministic probability flows. The proposed framework facilitates manipulation of the target distribution properties through a continuous control variable, which defines a certain aspect of the target domain. Notably, this approach does not rely on paired examples for training. To address identified challenges on maintaining the speech content consistent, we recommend a training strategy that incorporates chunk-based minibatch Optimal Transport couplings of data samples and noise. Comparing our unsupervised method with established baselines, we find competitive performance in tasks of reverberation and distortion manipulation. Despite encoutering limitations, the intriguing results obtained in this study underscore potential for further exploration.


\end{abstract}

\vspace{-5pt}
\section{Introduction}
\vspace{-5pt}
\label{sec:intro}
The search for data-driven methods that allow for controlled modification of audio signals has attracted considerable attention and research efforts throughout the past decade. 
The majority of proposed techniques are dependent on supervised learning, 
necessitating the availability of paired "input" and "target" audio samples for effective training. 
A prominent instance of this is speech enhancement, which has seen significant advancements in performance due to meticulously designed data processing pipelines and optimization strategies \cite{braun2022effect}.
However, the requirement for paired samples introduces substantial constraints, making it impractical in certain scenarios. Obtaining such data can be challenging or impossible due to the high cost and effort involved in producing or collecting it, limited access to specific real-world conditions, potential mismatches when using synthetic data, or difficulties in achieving precise time alignment between pairs. 
Consequently, unsupervised methods
offer a promising research avenue.
In the context of audio, there have been several contributions in this direction, employing techniques such as mixture-invariant training \cite{trinh2022unsupervised}, or various forms of generative adversarial networks \cite{kaneko2018cyclegan, fu2022metricgan, wright2023adversarial}.


Additionally, recent studies have demonstrated the effectiveness of diffusion models in unsupervised audio restoration and editing. 
A known approach consists of utilizing the generative priors from diffusion models to sample from posterior distributions \cite{moliner2023solving}. However, this framework requires exact knowledge of a forward degradation and is, thus, not readily applicable for general settings.
A work closely related to ours is by Popov et al. \cite{popov2023optimal}, who employed a bridge to transport mel-spectrograms for voice conversion and instrument timbre transfer.  
Similarly, Manor and Michaeli explored a related technique called "DDPM inversion" for unsupervised editing of mel-spectrograms \cite{manor2024zero}.





In this study, we explore a technique we term \emph{Gaussian Flow Bridges} (GFBs).
This method offers a general way to handle audio domain transfer tasks in an unsupervised manner.
GFBs address a transport problem between probability densities, known as the Schrödinger bridge problem \cite{de2021diffusion}, by applying two sequential deterministic processes or flows. 
The first process transforms an audio waveform into a latent vector within a Gaussian distribution, while the second changes this latent vector into a modified waveform, as shown in Figure 1.
GFBs enable many-to-many mappings within audio domains.
This approach aligns with concepts previously explored as "Dual Diffusion Implicit Bridges" \cite{su2022dual}, or "DDIM inversion" \cite{mokady2023null}.  
This paper applies this idea through the Flow Matching framework \cite{lipman2022flow}, hence the distinct terminology.

Unlike prior work \cite{popov2023optimal, manor2024zero}, our research focuses on the development of GFBs in the waveform domain.
This approach eliminates the need for a spectrogram inversion model, thereby simplifying the operational framework. 
A pivotal aspect of our study is addressing the complexities introduced by waveform representation in GFBs, particularly when maintaining content fidelity in speech signals.
These complexities often manifest as undesirable artifacts, including abrupt identity shifts or unintelligible speech.
Our work adheres to optimal transport principles and underscores the importance of 
linear transformation paths within the GFB framework, suggesting that maintaining linear trajectories is crucial for preserving content integrity.
To enhance the model performance while adhering to this principle, we introduce a training methodology that uses chunk-based minibatch optimal transport (OT) couplings. 

 The experiments outlined in Sec. \,\ref{sec:experiments} delve into two key areas: speech reverberation and distortion. While our methodology showcases promising results in these domains, it is important to emphasize the broader applicability of the discussed approach.

\vspace{-5pt}
\section{Background}
\vspace{-5pt}

\subsection{Continuous Normalizing Flows}
Continuous Normalizing Flows (CNFs) are designed to iteratively transport samples between two probability distributions, denoted as $\mathbf{x}_0 \sim q_0$ and $\mathbf{x}_1 \sim q_1$, across a defined \emph{time} interval $\tau \in [0,1]$. 
The underlying process is formalized with an Ordinary Differential Equation (ODE), characterized by a time-dependent vector field:
\begin{equation}
    \text{d}\mathbf{x}_\tau= u(\mathbf{x}_\tau, \tau)\text{d}\tau.
\end{equation}
With a specified vector field $u$, it becomes feasible to transport samples from $\mathbf{x}_0 \sim q_0$ to $\mathbf{x}_1 \sim q_1$ and vice versa by solving the ODE both in the \emph{forward} direction, where $\tau$ varies from 0 to 1, and in the \emph{backward} direction, where $\tau$ varies from 1 to 0.
Particularly, when one of these distributions is a tractable one, such as a Gaussian defined by $p_1 = \mathcal{N}(\mathbf{0}, \mathbf{I})$, CNFs function as generative models and exhibit notable parallels with diffusion models \cite{ho2020denoising}.


\vspace{-5pt}
\subsection{Conditional Flow Matching}



Several works \cite{lipman2022flow, liu2022flow} approximate the vector field $u(\mathbf{x}_\tau,\tau)$ with a deep neural network $v_\theta (\mathbf{x}_\tau, \tau)$.
This approximation is achieved by optimizing the Conditional Flow Matching objective:
\begin{equation}\label{eq:CFM}
\mathcal{L}_\text{CFM}=
\mathbb{E}_{\tau, q_0, q_1}
\lVert
v_\theta(\mathbf{x}_\tau,\tau)
-u(\mathbf{x}_\tau,\tau |  \mathbf{x}_0, \mathbf{x}_1)
\rVert^2
.
\end{equation}



As suggested in \cite{liu2022flow, lipman2022flow},
a valid strategy for designing the probability path is linear interpolation: $\mathbf{x}_\tau=(1-\tau)\mathbf{x}_0 + \tau\mathbf{x}_1$.
This choice yields a vector field with constant velocity over time:
    $u(\mathbf{x}_\tau, \tau | \mathbf{x}_0,  \mathbf{x}_1) = \mathbf{x}_1 -\mathbf{x}_0$,
which corresponds to straight linear trajectories.
When one of the distributions is a Gaussian, such linear trajectories represent optimal transport paths \cite{lipman2022flow}.


Conditional information, if available, can be used to direct the trajectories using a technique known as Classifier-Free Guidance \cite{ho2022classifier}, 
where a conditional vector $\mathbf{c}$ is added as an auxiliary input to the model. This allows the conditional influence to be modulated as
\begin{equation}
    \tilde{v}_\theta(\mathbf{x}_\tau, \mathbf{c}, \tau) = \gamma   v_\theta(\mathbf{x}_\tau, \mathbf{c}, \tau) + (1-\gamma)v_\theta(\mathbf{x}_\tau, \mathbf{c}=\emptyset, \tau),
\end{equation} 
where $v_\theta(\mathbf{x}_\tau, \mathbf{c}=\emptyset, \tau)$ implies that the conditioning vector $\mathbf{c}$ is not included, and the hyperparemeter $\gamma$ weights both model evaluations.

\vspace{-5pt}
\section{Methods}
\vspace{-5pt}
\begin{figure}
    \centering
    \includegraphics[]{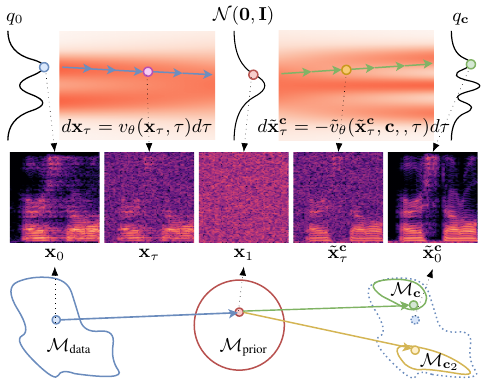}
    \vspace{-10pt}
    \caption{\textit{(Top) Illustration of a GFB in one-dimensional space. (Middle) A sequential display of spectrograms, showcasing the stages of audio signal transformation. (Bottom) Geometrical interpretation highlighting the mapping of data points through encoding and decoding within a Gaussian space.}}
    \vspace{-5pt}
    \label{fig:distributions}
\end{figure}

\subsection{Gaussian Flow Bridges} \label{sec:GFB}
The Gaussian Flow Bridges (GFB) method involves a two-step process 
using two deterministic flows evaluated in opposite directions, an encoder and a decoder. As represented in Fig. \ref{fig:distributions},  a starting sample \(\mathbf{x}_0 \in \mathbb{R}^n\) at \(\tau=0\) is first encoded using an unconditional vector field model \(v_\theta(\mathbf{x}_\tau, \tau)\) into a Gaussian distribution \(q_1 =\mathcal{N}(\mathbf{0}, \mathbf{I})\) at \(\tau=1\), producing a latent sample \(\mathbf{x}_1 \in \mathbb{R}^n\). This latent sample is then decoded with a conditional model \(\tilde{v}_\theta(\mathbf{x}_\tau, \mathbf{c}, \tau)\), yielding a modified sample \(\tilde{\mathbf{x}}_0^\mathbf{c} \in \mathbb{R}^n\). The condition \(\mathbf{c}\) is a continuous variable, enabling GFB to create a diverse range of outcomes based on \(\mathbf{c}\). 
 It is worth noting that the encoding and decoding processes, assuming ideally straight trajectories, are designed to be optimal transport paths between the different distributions and a Gaussian. 
 Nonetheless, this does not guarantee optimal displacement between the endpoints themselves.

The bottom of Fig \ref{fig:distributions} represents our interpretation of the GFB concept from a geometric perspective. The data points \(\mathbf{x}_0\) are described as part of a data manifold \(\mathcal{M}_\text{data} \subset \mathbb{R}^n\). Through the encoding, these points are mapped to a Gaussian noise space or hypersphere (\(\mathcal{M}_\text{prior}\)). During decoding, the conditional vector field guides these points to a specific subset of the original data manifold
\(\tilde{\mathbf{x}}_0^\mathbf{c} \in \mathcal{M}_\mathbf{c}\), which depends on \(\mathbf{c}\).
 We hypothesize that the optimal endpoints \(\mathbf{x}_0^\mathbf{c}\), those that reflect the attributes specified by \(\mathbf{c}\) while minimally altering unrelated features, should ideally lie in close Euclidean proximity to the initial sample \(\mathbf{x}_0\).
 This conjecture supports the notion that leveraging a Gaussian distribution as a bridge can facilitate a valid approximation.

\vspace{-5pt}
\subsection{Chunk-based minibatch optimal transport couplings}

As will be shown in the analysis Sec. \ref{sec:reverb_eval}, 
 we observe that GFBs struggle to preserve content that should be orthogonal to the conditioning variable $\mathbf{c}$.
 We hypothesize that the problem is linked to the trajectory curvature that appears during sampling.
As suggested in \cite{lee2023minimizing, liu2022flow, pooladian2023multisample}, such curvature
  arises from employing data-independent couplings during training.
  Specifically, when the pairs of data $\mathbf{x}_0$ and noise $\mathbf{x}_1$ are sampled independently, 
  the resulting training trajectories tend to intersect.
This intersection causes the model to approximate a suboptimal average of these paths, rather than identifying distinct, optimal paths individually \cite{liu2022flow}.
Such convergence towards an average trajectory deviates from the intended direct paths and can ptentially harm the efficacy and reliability of the GFB strategy.

With the goal of minimizing trajectory curvature, some works propose to assign the data/noise pairs during training using a minibatch optimal transport strategy \cite{pooladian2023multisample,tong2023improving} .
 According to their findings,  this approach effectively minimizes the trajectory curvature and reduces the variance of gradients during training \cite{pooladian2023multisample}.
However, we realize that this strategy does by default not scale well for data of very high dimensionality, such as audio waveforms, which are typically sampled at high rates.
As the dimensionality increases, the number of possible data configurations grows exponentially, necessitating larger minibatch sizes for effective coverage.
This leads to increased computational and memory requirements, 
potentially causing bottlenecks in our training pipeline.

To adapt this approach for practical use with audio data, we propose a redefinition of minibatches into smaller chunks. 
Our observation is that speech is characterized by high information density, with significant data concentration occurring within localized time windows of just a few milliseconds. 
The original minibatch $\{ \mathbf{x}_0^{(i)} \in \mathbb{R}^N\}_{i=0}^{B}$
is partitioned into a minibatch of chunks $\{ \mathbf{x}_0^{(k)} \in \mathbb{R}^{N_\text{c}}\}_{k=0}^{B_\text{c}}$, where the chunk length is notably smaller than the original ($N_\text{c} \ll N$ ), and the minibatch size increases to $B_\text{c}=B\frac{N}{N_\text{c}}$.

The following step is to compute an optimal transport coupling between the chunked minibatch and an equal-sized set of noise samples. 
 A matrix of pairwise L2 distances between all the elements in the two minibatches is computed, and the corresponding pairs are assined with an optimal transport solver. We use off-the-shelf solvers from the \emph{Python Optimal Transport } library \cite{flamary2021pot}, in particular, when using $N_\text{c}=512$,  we use $\mathrm{ot.emd}$, an exact optimal transport solver. 
 When experimenting with smaller $N_c$, we instead opted for \emph{ot.sinkhorn}, an entropy-regularized solver that provides approximate OT solutions at a lower computational cost.
 After, the coupled pairs are reshaped to their original dimensions and the CFM objective (Eq. \ref{eq:CFM}) is computed.
The training loop employing the chunk-based minibatch OT methodology is detailed in Algorithm 1.

\begin{algorithm}[t]
\caption{Training with chunk-based minibatch OT}
\label{alg}
\begin{algorithmic}
\For{each training iteration}
\State Sample minibatches $\{\mathbf{x}_0^{(i)}\}_{i=0}^B \sim q_0$  and $\{\mathbf{x}_1^{(j)}\}_{j=0}^{B} \sim q_1 $
\State Split ($\mathbf{x}_0$, $\mathbf{x}_1$) into chunks
\State Compute $\mathbf{C}_{ij} = \lVert \mathbf{x}_0^{(i)} - \mathbf{x}_1^{(j)} \rVert_2^2$
\State Solve OT for $\mathbf{C}_{ij}$, get coupling ($\overline{\mathbf{x}}_0$, $\overline{\mathbf{x}}_1$)
\State Reshape ($\overline{\mathbf{x}}_0$, $\overline{\mathbf{x}}_1$) to the original length

\State $\mathbf{x}_\tau \leftarrow (1-\tau)\overline{\mathbf{x}}_0 + \tau\overline{\mathbf{x}}_1$ , $\tau \sim \mathcal{U}(0,1)$

\State $\mathcal{L}_\text{CFM}(\theta) \leftarrow 
\mathbb{E}_{\tau,  q_0,  q_1 }
\lVert 
v_\theta(\mathbf{x}_\tau, \tau) - (\overline{\mathbf{x}}_1 -\overline{\mathbf{x}}_0)
\rVert^2_2
$
\State $\theta \leftarrow  \mathrm{update} (\theta, \nabla_\theta\mathcal{L}_\text{CFM}(\theta)$)
\EndFor

\end{algorithmic}
\end{algorithm}

\vspace{-5pt}
\section{Experiments} \label{sec:experiments}
\vspace{-5pt}

We conducted experiments in two areas, namely \emph{speech reverberation} and \emph{distortion}. In \emph{speech reverberation}, we focused on the task of acoustics transfer, which extends beyond the dereverberation task to modifying the characteristics of reverberation.
Such a controllable approach holds potential for a variety of applications in augmented and virtual reality \cite{chen2022vam}, where the ability to modify audio signals to match expected acoustics can significantly enhance listener experience.
We design a GFB where the initial distribution $p_\text{data}$ contains speech with undetermined acoustic properties and the terminal distribution $p_\textbf{c}$ comprises speech signals with a specified acoustic condition $\textbf{c}$. We experiment with two reverberation descriptors: reverberation time (T$_{60}$) and clarity (C$_{50}$).

For \emph{distortion}, we explore our method's ability to handle non-linear effects, with our experiments focusing on \emph{speech clipping}.
Here, the GFB is trained with both clipped and clean speech signals, and the goal is to transform initial samples to a specific Signal-to-Distortion Ratio (SDR).
Additionally, we provide qualitative insights into \emph{guitar distortion} manipulation in the companion webpage\footnote{Code and examples available at \href{https://github.com/microsoft/GFB-audio-control}{github.com/microsoft/GFB-audio-control}}.




\vspace{-5pt}
\subsection{Experimental setup}

As training data, we used studio quality speech samples from VCTK \cite{yamagishi2019cstr}.
For our reverberation experiments, we convolved the speech recordings with single-channel room impulse responses (RIRs), collected by combining several public datasets \cite{prawda2022calibrating, szoke2019building, Steward2010database, murphy2010openair, Eaton2015ACE}, using RIRs with T$_{60}$ values ranging from 0 to 1s.
For the clipping experiment, the training speech samples were clipped at different SDR levels. 
During the training, we also include, with a probability of 10\%, clean speech samples.
The reverberation descriptors T$_{60}$ and C$_{50}$, and the SDR in the case of declipping, are estimated and concatenated into a conditioning vector $\mathbf{c}$.
All signals are resampled to 16\,kHz and are randomly cropped to a segment size of 4.09\,s.

In our experiments, we use a backbone architecture $v_\theta$ based on the Short-Time Fourier Transform (STFT). 
A forward and an inverse STFT are applied wrapping trainable neural network layers in a similar way as in \cite{moliner2023solving}.
The complex-valued spectrograms are processed as double-real signals,
stacking the real and imaginary parts into the channel dimension.
The architecture is a U-Net with roughly 44\,M trainable parameters, mainly consisting of 2-Dimensional convolutional layers, 
The conditioning vector $\mathbf{c}$, alongside with the time variable $\tau$, is fed into the neural network through feature modulations.
During training, the conditioning vector $\mathbf{c}$ is randomly dropped with a probability of 20\%, to allow unconditional sampling and the use of Classifier-Free Guidance.
All models compared in the experiments are trained for 300k iterations using the Adam optimizer, with a learning rate of $10^{-4}$,  and a batch size of 8.









\vspace{-5pt}
\subsection{Coupling configurations and trajectory curvature analysis} \label{sec:curvature_analysis}

Our investigation begins by examining the influence of the training couplings on the sampling trajectories. Following the methodologies described in \cite{lee2023minimizing, liu2022flow}, we utilize a surrogate metric to analyze trajectory curvature
$C(t)=
(\mathbf{x}_1-\mathbf{x}_0) - \frac{\partial\mathbf{x}_t}{\partial t}$, which compares the local slope at every time \(\tau\) with the total displacement. Ideally, if the paths were  completely straight, this metric should yield a value of 0.







\begin{figure}
    \centering
    \includegraphics[width=1\columnwidth]{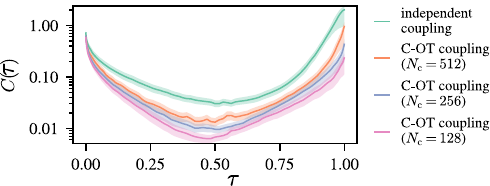}
    \vspace{-15pt}
    \caption{\textit{Averaged trajectory curvature with respect to time $\tau$ when different coupling strategies are used. The shaded area represents the 25\% and 75\% percentiles.}}
    \vspace{-10pt}
    \label{fig:curvature}
    
\end{figure}

Figure \ref{fig:curvature} displays the distribution of $C(\tau)$ values for different timesteps during the forward sampling process. These trajectories begin from each example in the test set at $\tau=0$ and progress toward a Gaussian distribution at $\tau=1$.
We compare the results obtained with a model trained on the reverberant speech dataset with the default training setup (independent couplint) against other models trained with the proposed chunked minibatch OT (C-OT) couplings. 
For the latter, we study the effect of the chunk length $N_c$ which, assuming a fixed sample length $N$ and batch size $B$, affects directly the chunked minibatch size $B_\text{c}=B\frac{N}{N_\text{c}}$.
Three different chunk lengths $N_ \text{c}$ are considered: 512, 256, and 128 samples; which correspond to 32, 16 and 8\;ms.

The results reveal that the use of C-OT couplings significantly reduces the observed curvature, with a consistent reduction when the chunk size \(N_c\) is decreased, showcasing the effect of the C-OT couplings. 
It can also be observed that all configurations show lower curvature values around the midpoint of the process ($t \approx 0.5$), but these values notably increase toward the extremes, specially at $t \approx 1$.
 This observed behavior inspires the adoption of a discretization scheme based on a raised cosine schedule that prioritizes smaller time steps at the extremes:
$   \tau_{i<T} =0.5+0.5\cos(\pi i/T + \pi)).
$
 We used this schedule, with $T=25$ steps,  in the rest of experiments.
 Although not the primary focus of this study, the observed lower curvature of C-OT couplings indicate a potential for more efficient sampling compared to conventional flow or diffusion-based models. 







\vspace{-5pt}
\subsection{Speech reverberation evaluation}\label{sec:reverb_eval}

 

Our investigation focuses on assessing the performance of different models in reverberation control, emphasizing the trade-off between two aspects: \emph{acoustics accuracy} and \emph{speech content consistency}.
Acoustics accuracy assesses the model's capability to recreate speech aligned with predetermined acoustic features, particularly T60 and C50. 
To quantify the models' fidelity in reproducing the target acoustic characteristics, we utilize a blind acoustic parameter estimator \cite{Gamper2018blind}.
We calculate the mean absolute error between the model-predicted T60 and C50 values and their actual measurements. 


When analyzing speech content consistency, we assess the model's ability to retain the original speech content. This involves addressing two critical issues: alterations in speaker identity and potential loss of intelligibility.
To adress the first issue, we measure the cosine similarity between embeddings derived from the speaker recognition model \cite{wang2022efficienttdnn}, we refer to this metric as Speaker Recognition Cosine Similarity (SR-CS).
With respect to the second, we compare Automatic Speech Recognition transcripts before and after the GFB. 
We use the "small" version of Whisper \cite{radford2023robust} and report the Word Error Rate (WER) with the original sample's transcription as a benchmark.  
We assume these two models to be robust and approximately invariant to acoustics.


We use a test set conforming 20 minutes of 4-s length studio quality speech examples from DAPS \cite{mysore2014can}, a different dataset than the one used for training.  
These examples are convolved with a set real RIRs containing uniformly balanced T60 values ranging from 0 to 1s, and C50 values ranging from 0 to 25 dB. We use 320 RIRs extracted from datasets not used during training \cite{traer2016statistics, carlo2021dechorate, kinoshita2013reverb, nielsen2014SMARD}.
All the examples in the test set are transformed to 8 different endpoints using the proposed GFB, each of them corresponding to a distinct conditioning setting with specific T60 and C50 values.

\begin{figure}
    \centering
    \includegraphics[width=\columnwidth]{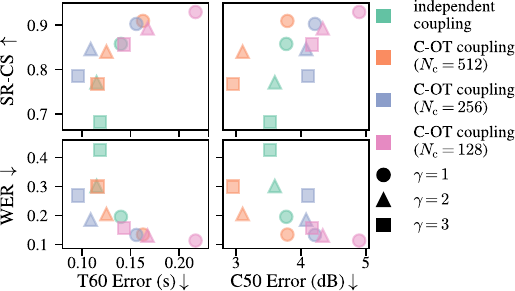}
    
    \vspace{-6pt}
    
\caption{\textit{Scatter plots illustrating the trade-offs between SR-CS and WER versus T$_{60}$ and C$_{50}$ errors for models conditioned on specific acoustic features. Points represent aggregated test set results, highlighting the effects of chunk length (N$_\text{c}$) and CFG scale ($\gamma$). }}
\vspace{-5pt}
    \label{fig:trends}
\end{figure}

In Figure \ref{fig:trends}, we provide a detailed analysis reflecting the average SR-CS and WER in relation to both T$_{60}$ and C$_{50}$ errors. These scatter plots are generated from the averaged results across the test set. The figure illustrates the outcomes for various models trained using different C-OT couplings, wherein the chunk length \(N_\text{c}\) varies. Additionally, each model's behavior concerning the Classifier-Free Guidance scaling parameter $\gamma$ is examined.
Our observations reveal that, for smaller $N_\text{c}$, the speech consistency gets improved, but usually at the cost of reduced acoustic accuracy. 
In addition, the parameter $\gamma$ reflects a notable trade-off, as increasing this value allows for more precise adjustments in acoustics characteristics at the expense of introducing artifacts that compromise speech consistency.

\begin{figure}
    \centering
    \includegraphics[width=0.96\columnwidth]{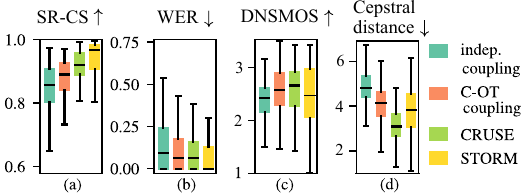}

    \vspace{-10pt}
    \caption{\textit{Objective evaluation on speech dereverberation. }}
    \label{fig:results_dereverb}
\end{figure}

Additionally, we assess the proposed method performance for dereverberation.
We utilize the speech consistency metrics \emph{SR-CS} and \emph{WER}, the MOS prediction metric \emph{DNSMOS} \cite{reddy2021dnsmos}, and the cepstral distance \cite{kitawaki1988objective}.
In our evaluation, two state-of-the-art baselines are included, CRUSE \cite{braun2022effect} and STORM \cite{lemercier2023storm}. Unlike the proposed method, these baselines were trained using paired data.
Results are shown on a subset of the test set, specifically 160 utterances with T60 values falling within the range of 0.5 to 1. 
The conditioning parameters for the diffusion bridge are set to T60=0.1s and C50=20dB, a dry but not anechoic specification.
 We conduct a comparative analysis between the model trained with independent coupling and the one trained with C-OT coupling, employing $N_\text{c} = 512$ and $\gamma=1$.
The results are presented in Figure \ref{fig:results_dereverb}.
Although none of the compared versions of the proposed method surpass the baseline performance, the results of the C-OT method notably converge closely.
It should be noted that the proposed method and the baselines were trained using different data, thus the results depend on the models' generalization capabilities.

\vspace{-5pt}
\subsection{Declipping evaluation}



\begin{figure}
    \centering
    \includegraphics[width=0.96\columnwidth]{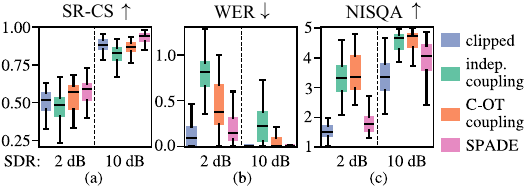}
    \vspace{-10pt}
    \caption{\textit{Objective evaluation on speech declipping.   }}
    \vspace{-10pt}
    \label{fig:declipping}
\end{figure}

The performance of both versions of our method in the task of speech declipping is compared against the clipped speech and SPADE \cite{kitic2015sparsity}, a popular sparsity-based declipping baseline.
In Figure \ref{fig:declipping}, we report three objective metrics: SR-CS and WER, as introduced in Section \ref{sec:reverb_eval}, and NISQA \cite{mittag2021nisqa}, a MOS prediction model that is known to correlate well with declipping performance. 
The results show a strong improvement of the proposed method against SPADE in terms of NISQA. However, in terms of WER and SR-CS, GFB does not reach the same consistency scores of SPADE, and neither of the clipped speech.
We also notice that the usage of C-OT couplings is critical at reducing the WER in this setting.




\vspace{-5pt}
\section{Conclusion}
\vspace{-5pt}
This paper studied the application of GFBs for unsupervised audio domain transfer, with experiments on reverberation and distortion control.
 The experiments show that, in the majority of cases, GFBs effectively manage to alter an audio effect characteristic while preserving the content integrity, a notable achievement considering it was not specifically trained for this task. 
  Furthermore, the method exhibits the ability to generalize to unseen speakers and acoustic conditions. 
  Qualitative assessments indicate that GFBs yield results free from typical artifacts seen in speech reverberation and declipping. However, occasional inconsistencies in speech content and speaker identity are observed, 
  posing a significant challenge for the method's potential applications.
Nonetheless, the performance of GFBs shows promising progress, paving the way for further enhancements and applications in diverse tasks and domains.

\vfill\pagebreak

%
\bibliographystyle{IEEEbib}
\bibliography{strings,references}

\end{document}